\documentstyle[12pt]{article}

\newcommand{\dis}{\displaystyle}

\newcommand{\extraspace}{\addtolength{\abovedisplayskip}{2mm}
                        \addtolength{\belowdisplayskip}{2mm}
                        \addtolength{\abovedisplayshortskip}{2mm}
                        \addtolength{\belowdisplayshortskip}{2mm}}
\newcommand{\be}{\begin{equation}\extraspace}

\newcommand{\ee}{\end{equation}}

\newcommand{\bea}{\begin{eqnarray}\extraspace}
\newcommand{\beastar}{\begin{eqnarray*}\extraspace}
\newcommand{\eea}{\end{eqnarray}}
\newcommand{\eeastar}{\end{eqnarray*}}

\newcommand{\nonu}{\nonumber \\[2mm]}

\newcommand{\strutje}{\rule[-1.5mm]{0mm}{5mm}}

\newcommand{\xb}{\bar{x}}

\newcommand{\ef}{{\rm eff}}
\newcommand{\ind}{{\rm ind}}

\newcommand{\del}{\partial}
\newcommand{\delb}{\bar{\partial}}

\def\b{\beta}

\def\d{\delta}
\def\e{\epsilon}                % Also, \varepsilon
\def\f{\phi}                    %       \varphi
\def\g{\gamma}

\def\l{\lambda}

\def\p{\pi}                     % Also, \varpi
\def\G{\Gamma}

\def\L{\Lambda}

\newcommand{\np}{Nucl.\ Phys.\ }

\newcommand{\cmp}{Comm.\ Math.\ Phys.\ }
\newcommand{\pl}{Phys.\ Lett.\ }

\catcode`\@=11
\def\@afterindentfalse{\let\if@afterindent=\iftrue}
\catcode`\@=12

\begin{document}
\font  \biggbold=cmbx10 scaled\magstep2
\font  \bigbold=cmbx10 at 12.5pt
\font  \bigreg=cmr10 at 12pt

\baselineskip = 15pt

\noindent Nov. 1991 \hfill ITP-SB-91-50\\
$\strutje$ \hfill  LBL-31474, UCB-PTH-91/66\\

\vspace{4mm}

\begin{center}
{\large QUANTUM}\ {\Large $W_3$}\ {\large GRAVITY%
\footnote{Contribution to the Trieste Summerschool on
{\it High Energy Physics and Cosmology}, Aug.\ 1991.}%
\footnote{Work supported in part by grant NSF-91-08054.}%
\footnote{This work was supported in part by the Director,
Office of Energy Research, Office of High Energy and Nuclear Physics,
Division of High Energy Physics of the U.S. Department of Energy
under Contract DE-AC03-76SF00098 and in part by the National Science
Foundation under grant PHY90-21139.} }\\

\vspace{1cm}

{\bf \centerline{K. Schoutens${}^1$, A. Sevrin${}^2$ and
P. van Nieuwenhuizen${}^{3,1}$}}
\vskip .4cm
{\baselineskip = 12pt
\centerline{\sl{1. Institute for Theoretical Physics}}
\centerline{\sl{State University of New York at Stony Brook}}
\centerline{\sl{Stony Brook, NY 11794-3840, U.S.A.}}}
\vskip .3cm
{\baselineskip = 12pt
\centerline{\sl{2. Department of Physics}}
\centerline{\sl{University of California at Berkeley}}
\centerline{\sl{and}}
\centerline{\sl{Theoretical Physics group}}
\centerline{\sl{Lawrence berkeley Laboratory}}
\centerline{\sl{Berkeley, CA 94720, U.S.A.}}}
\vskip .3cm
{\baselineskip = 12pt
\centerline{\sl{3. Theory Division, CERN}}
\centerline{\sl{CH-1211 Geneva 23}}
\centerline{\sl{Switzerland}}}
\end{center}

\vskip 1.4cm
\centerline{\bf Abstract}
\vskip 0.5cm
\baselineskip=14pt

\noindent We briefly review some results in the theory of quantum
$W_3$ gravity in the chiral gauge. We compare them
with similar results in the analogous but simpler cases of
$d=2$ induced gauge theories and $d=2$ induced gravity.

\newpage
\baselineskip=18pt

Gravity in two dimensions has the exceptional property that it allows for
higher spin extensions: the $W$ gravity theories
(reviews of the classical formulations of these theories can be found
in \cite{review}). The basic structure behind these theories are
$W$ algebras, which are higher-spin extensions of the Virasoro algebra.

In a recent project, we studied the quantum theory for the particular
theory of chiral gauge $W_3$ gravity in considerable detail. Extensive
accounts of this work can be found in the research papers
\cite{us3}-\cite{us5} and in the review papers \cite{veltman,stony}.
In our contribution to these proceedings we will recall the main results that
were obtained and give some comments; for detailed derivations and proofs we
refer to the papers just cited.

The first step in the construction of classical $W_3$ gravity was made
by Hull \cite{hull}, who constructed a classical theory of
free scalars coupled to $W_3$ gravity in the chiral lightcone gauge.
Subsequently, more covariant
formulations were given in \cite{us1,us2}. A similar program for classical
$w_\infty$ gravity was carried out in \cite{winfty}.

These developments were purely classical. In $W$ gravities, the
classical theory (in the gauge sector) is in some sense trivial as
there are as many gauge field components as local symmetries. This is in fact
a phenomenon which occurs often in two-dimensional conformally invariant
theories: a classical theory with local invariances has as many invariances
as there are gauge field components. Classically, the gauge sector is then
pure gauge (up to moduli). However, at the quantum level, some of the
symmetries may become anomalous and thereby some gauge degrees of freedom
may become propagating.

Typical examples for this `anomalous quantization' are induced gauge
theories and the theory of $d=2$ induced gravity. Let us focuss on
the latter and consider free massless scalar fields coupled to gravity.
The gauge multiplet consists of four components: the zweibein. The theory
has also four gauge invariances: diffeomorphisms, local Lorentz and Weyl
invariance. Quantizing the theory, one finds that one of these symmetries
becomes anomalous. At that
point two strategies are available: either one cancels the anomaly
through a suitable choice of background (which would be 26 scalar fields
in this example) so that the over-all coefficient of the integrated
anomaly vanishes (the so-called {\it critical}\ approach), or one learns
how to solve the quantum theory in the presence of propagating gauge degrees
of freedom (the {\it non-critical}\ approach). In the latter case,
integrating out the matter fields will result in a quantum action for the
gauge fields: the induced action. The action obtained from there by performing
a path-integral over the (propagating) gauge fields is called the effective
action.

$W_3$ gravity is another example of a theory with a non-trivial
quantum induced action. (The same is of course true for $W$ gravity
theories in general.) The structure of the anomalies in the chiral
light-cone gauge has been studied in \cite{matsuo}-\cite{hullpal},
some results departing from the covariant formulation were obtained
in \cite{anna}. Contrary to what
happens for induced gauge theories or for pure gravity, one finds that the
induced action is not proportional to the central charge $c$. This
phenomenon, which is due to the non-linearities in the $W_3$ algebra,
leads to considerable complications which we will address below.

\vspace{6mm}

Since $d=2$ induced gauge theories and the theories of $d=2$ gravity
and $W$ gravity are similar, we think it worthwhile to discuss
them in parallel. We first focuss on induced gauge theories in two
dimensions (see for example \cite{poly}). Let us suppose that we are given a
matter system which has rigid symmetries that generate an affine Kac-Moody
algebra. In the language of current algebra this would mean that we
can define currents $J_a(x,\xb)$, which are conserved on-shell
($\delb J_a(x,\xb)=0$)%
\footnote{ We work in two-dimensional euclidean space with complex
           coordinates denoted by $x$, $\xb$}
and which satisfy the following Operator Product Expansions (OPE's)
\be
J_a (x) J_b (y) = - \frac{k}{2} g_{ab} (x-y)^{-2} + (x-y)^{-1} f_{ab}{}^c
J_c (y) + \cdots \ .
\label{kmope}
\ee
The c-number $k$ is an integer and is called the level of the affine algebra.
A typical example would be a theory of free fermions, taking values in the
adjoint representation of a Lie algebra $g$, which would give rise to the
current algebra (\ref{kmope}) for the affine algebra $g^{(1)}$. In that case,
$k$ would be equal to $\tilde{h}$, which is the dual Coxeter number of the
Lie algebra $g$.

We can now consider the coupling of this matter system to gauge fields
$A^a(x,\xb)$. This will promote the rigid affine symmetries to local
gauge symmetries. The gauge-field sector of the coupled theory is
trivial in the sense described above: both in the left and the right
moving sectors we would have $\dim g$ gauge-field components and
$\dim g$ local symmetries. However, at the quantum level this balance
of degrees of freedom can work out differently due to the occurence of
anomalies. One can study this phenomenon by considering the so-called
quantum induced action $\G_\ind[A]$ for the gauge fields, which is obtained
by integrating out the matter degrees of freedom in a path-integral.
In the chiral gauge, where we keep only one chirality of the gauge-fields
(say, the left-movers), we have the following relation%
\footnote{ We normalize such that if $[T_a, T_b]=f_{ab}{}^c T_c$
    then $f_{ac}{}^d f_{bd}{}^c = - \tilde{h} g_{ab}$, where $\tilde{h}$
    is the dual Coxeter number. In a representation $R$ we have
    $tr(T_aT_b)=-x g_{ab}$, where $x$ is the index of the representation
    $(x=\tilde{h}$ for the adjoint representation).}
\be
e^{\dis -\G_\ind[A]} = \, \langle \,
 \exp - \frac{1}{\p x} \int d^2 x \, tr \left\{ J(x) A(x) \right\}
                \, \rangle \, .
\label{indacI}
\ee

An important observation is that the leading term in the induced
action $\G_\ind [A]$ is a quadratic term, which can be viewed as a kinetic
term for the gauge fields $A^a(x,\xb)$. Roughly speaking, one could
say that the presence of quantized matter fields `induces' propagation
of the gauge fields in the theory. Of course, this phenomenon violates
what one would expect on the basis of gauge invariance, which, as we saw,
precisely balanced the gauge-field degrees of freedom at the classical
level. Indeed, the kinetic term in the induced action is closely
related to an anomaly in the gauge invariance, the strength of which
is proportional to the level $k$.

The induced action for chiral gauge fields can formally be written as
\be
\G_\ind [A] =
\frac{k}{2\p x} \int d^2 x \; tr \left\{ A \sum_{n\geq 0}
 \frac{1}{n+2} \left(
\left[ \frac{1}{\bar{\del}} A, \cdot \right] \right)^n \frac{\del}
{\bar{\del}} A
\right\} .
\label{indacII}
\ee
We now write
\be
\G_\ind[A] = - k \, \G_{WZW}[A] \, ,
\label{nogeen}
\ee
which defines a $k$-independent reference functional $\G_{WZW}[A]$.
This functional is related to the well-known Wess-Zumino-Witten action
as follows \cite{polwi,orlando}. If we parametrize $A$ as
$A\equiv \bar{\del} g g^{-1}$
we find that $\G_{WZW}[A(g)]$ is precisely equal to the WZW action
with $k=1$. Thus we see that the change of variables from $A$ to
$g$ makes it possible to write the induced action in a form which
avoids non-localities. We will later find the analogous `local'
formulations of the induced actions for gravity and $W_3$ gravity.

We would now like to `quantize' the gauge-field degrees of freedom,
which were up to now treated as external fields. To this end, we
consider the following functional integral
\be
e^{\dis -W[t]}=\int {\cal D} A \,
e^{\dis -\G_\ind[A]+\frac{k}{2\p x}\int d^2 x \;  tr \{ tA \} } \, .
\label{pathint}
\ee
In here, $W[t]$ is the generating functional of connected diagrams with
propagating $A$ fields. The Legendre transform of this functional
yields the generating functional for 1PI diagrams, which is usually
called the effective action.
It can be argued \cite{stony} that
the effective action takes the following form
\be
\G_{\rm eff}[A]= - (k+2\tilde{h})\,
   \G_{WZW}\left[ \frac{k}{k+\tilde{h}} A \right] \, .
\label{effacI}
\ee
According to this, the effective action is again given in terms of
the reference functional $\G_{WZW}$, which is now `dressed' through
finite multiplicative renormalizations. We remark that the quantity
$1/k$ plays the role of `Planck's constant' for the quantization
of the gauge-fields, which is clear from the fact that the induced
action is proportional to $k$. As such, the loop-expansion of the
path-integral (\ref{pathint}) is equivalent to a $1/k$ expansion.
[The result \ref{effacI} has been checked through order $1/k$
by an explicit 1-loop calculation.]
For $k$ large, the effective action reduces to the `classical'
induced action (\ref{nogeen}).

\vspace{6mm}

Before turning to $W_3$ gravity we will first look at induced
pure gravity, which shares many characteristics with $W$ gravities
but avoids the problems associated with the non-linearities inherent
to $W$ symmetries.

By $d=2$ gravity one always means conformal gravity, and the matter
systems that can be coupled are thus Conformal Field Theories (CFT's).
As is well-known, these give rise to conserved currents $T(x)$ and
$\overline T(\xb)$, with short distance expansion
\be
T(x)T(y)= \frac{c}{2}(x-y)^{-4} +2(x-y)^{-2}T(y) +(x-y)^{-1}\del T(y)
          +\cdots \ ,
\ee
where $c$ denotes the central charge. (This relation is equivalent to the
Virasoro algebra.) One can for example think about
free scalar fields $\f^i$, $i=1,2,\cdots,n$, which would give a central
charge $c=n$.

In the chiral gauge, the only non-trivial component of the metric field
is $h_{--}$. Coupling this to a scalar field theory gives the classical
action
\be
S[\f^i , h]=\frac{1}{2\p}\int d^2 x \;
  \left( \del\f^i\bar{\del}\f^i+2hT\right) ,
\ee
where
\be
T=-\frac 1 2 \del\f^i\del\f^i \,.
\ee
Classically, this action has a local gauge invariance
\bea
\d \f &=& \e\del\f \nonu
\d h  &=& \bar{\del}\e+\e\del h-\del\e h \ ,
\eea
which balances the degree of freedom represented by $h$.

The induced action $\G_\ind[h]$ for gravity coupled to a general
CFT is defined by
\be
e^{\dis -\G_\ind [h]} = \langle \, \exp - \frac{1}{\p} \int d^2 x \;
   T(x) h(x) \, \rangle \,.
\label{indach}
\ee
The induced action can be given in a closed (albeit non-local) form
\be
\G_\ind[h] = \frac{c}{12} \G_L[h]
\ee
with
\be
\G_L[h] =
    \frac{1}{2 \p} \int d^2 x \; \del^2 h
    \frac{1}{\delb} \frac{1}{1-h \frac{\del}{\delb}}
    \frac{1}{\del} \del^2 h  \,.
\label{indacg}
\ee
In the scalar field theory, one can think of this result as the
sum of an infinite set of contributions, that correspond to
Feynman diagrams containing a $\f$ loop and $2,3,\cdots$ external
$h$ lines. The expression (\ref{indacg}) is easily recognized as the
reduction to the chiral gauge of the covariant result
$\int \sqrt{g} R \, \Box^{-1} R$. The local expression
for this action was first
given by Polyakov in \cite{pol}. It is found by introducing a coordinate
$f$ which is related to $h$ through $h=\delb f/(\del f)$. In terms
of $f$, the induced action reads
\be
\G_L[h(f)] =
 \frac{1}{2\pi} \int d^2 x \, \frac{\delb f}{\del f}\del^2
 \left(\ln\del f\right) .
\label{polact}
\ee

It is important for us to understand the systematics of the change
of variables from $h$ to $f$, since we will need to make a similar
but much more complicated step in the case of $W_3$ gravity.
It has been found that the introduction of $f$ follows naturally
if one approaches the induced action from the point of view of its
`hidden' $Sl(2,{\bf R})$ invariance. It was shown in \cite{pol2}
(see also \cite{bo,as}),
that the Ward Identities and the induced action for induced gravity
can be obtained from the similar quantities in the induced gauge
theory for $Sl(2,{\bf R})$, by applying a reduction procedure.
We will here not go into the details of this procedure, but we
would like to stress once more that it introduces the variable
$f$ in a natural way.

In the quantization of the $h$ field, the quantity $1/c$ plays the
role of Planck's constant. We define
\be
e^{\dis -W[t]} = \int {\cal D} h \,
e^{\dis -\G_\ind [h]+\frac{1}{\p} \int d^2 x \; h t }
\label{pathintII}
\ee
and we define $\G_\ef[h]$ as the Legendre transform of the
generating functional $W[t]$. The saddle-point approximation to the
path-integral (\ref{pathintII}) gives the leading terms in
$\G_\ef[h]$, which simply coincide with the induced action.
The 1-loop corrections to the saddle-point result can be computed by
standard determinant techniques, as was demonstrated by Polyakov
(unpublished, see \cite{alzamo}). The relevant
determinant is the determinant of the second derivative of the
induced action $\G_\ind[h]$ with respect to $h(x)$ and $h(y)$,
evaluated at the saddle-point. The 1-loop corrections lead to order
$1/c$ corrections in $\G_{\rm eff}[h]$. By using the results of
KPZ \cite{kpz}, one can then extend these perturbative results to an
exact expression for the effective action, which reads
\be
\G_{\rm eff}[h] = \frac{k}{2} \, \G_L\left[ \frac{k+2}{k} h \right] ,
\ee
where $k$ is given by
\be
k = - \frac{1}{12} \left( 25-c+ \sqrt{(c-1)(c-25)} \right) - 2 \,.
\ee

\vspace{6mm}

Let us now turn to a discussion of the case of induced
$W_3$ gravity. We start by choosing an appropriate matter
system, which should be a CFT that has so-called $W_3$ symmetry.
This means that it posseses conserved (chiral) currents $T(x)$
and $W(x)$ (plus their antichiral counterparts), which obey the
following OPE's
\bea
T(x) T(y) &=& \frac{c}{2} (x-y)^{-4} + 2(x-y)^{-2} T(y) + (x-y)^{-1} \del T(y)
+\cdots
\nonu
T(x) W(y) &=& 3 (x-y)^{-2} W(y) + (x-y)^{-1} \del W(y)
+\cdots
\nonu
W(x) W(y) &=& \frac{c}{3} (x-y)^{-6} + 2 (x-y)^{-4} T(y) + (x-y)^{-3} \del
T(y)
\nonu
&& + (x-y)^{-2} \left[ 2 \b \L (y) + \frac{3}{10} \del^2 T(y) \right]
\nonu
&& + (x-y)^{-1} \left[ \b \del \L (y) + \frac{1}{15} \del^3 T(y) \right]
   +\cdots ,
\label{WIII}
\eea
where
\be
\L (x) = (TT) (x) - \frac{3}{10} \del^2 T(x)
\label{lamb}
\ee
and
\be
\b = \frac{16}{22+5c} \, .
\label{bet}
\ee
For the purpose of extracting a string-theory interpretation,
one might wish to represent the matter system in terms of
scalar fields in the presence of so-called background charges
(these are needed if the number $n$ of scalar fields is different
from 2). We will later make some comments on this.

In analogy to our treatment of induced gauge theories and pure
gravity, we can now define the induced action of chiral $W_3$ gravity
by the following formula
\be
e^{\dis -\G_\ind[h,b]} = \langle \, \exp - \frac{1}{\p} \int d^2 x \left[
h (x) T(x) + b(x) W(x) \right] \, \rangle \, .
\label{indacw}
\ee

In contrast with what we found for pure gravity, we find here
that the induced action is not simply proportional to the central
charge $c$. Rather, it can be given by a $1/c$ expansion
\be
\G_\ind [h,b]= \frac{1}{12} \, \sum_{i\geq 0}c^{1-i}\, \G^{(i)}[h,b].
\label{cexp}
\ee
This phenomenon is due to the non-linearities in the $W_3$ algebra.
More precisely, it arises because the expectation value of the
quantity $\beta \L(z)$ has a non-trivial $c$-dependence.

Explicit expressions for all the terms $\G^{(i)}[h,b]$ in the $1/c$
expansion of the induced action seem to be beyond reach. However,
it is possible to explicitly evaluate the leading term $\G^{(0)}[h,b]$,
which dominates the induced action in the limit $c \rightarrow \pm\infty$.
As in the case of pure gravity, this is done by reducing an induced
gauge theory, which for the case at hand is based on $Sl(3,{\bf R})$.

To explain the reduction procedure, we first mention that the action
$\G^{(0)}[h,b]$ is completely characterized by the following Ward
Identities, which take the form of local differential equations
\bea
\delb u &=& D_1 \, h +
       \left[ \frac{1}{10}\, v \, \del + \frac{1}{15}
 \,(\del v) \right] b \,,
\nonu
\delb v &=& \left[ 3 \, v \, \del + (\del v) \right] h
             + D_2 \, b,
\label{wi}
\eea
where
\be
u(x) = \pi  \frac{\d \G^{(0)}[h,b]}{\d h(x)} \,, \qquad
v(x) = 30 \pi  \frac{\d \G^{(0)}[h,b]}{\d b(x)}
\label{henb}
\ee
and $D_1$ and $D_2$ are the 3rd and 5th order Gelfand-Dickey
operators given by (the primes denote $\del$)
\bea
D_1 &=& \del^3 + 2 \, u \del + u^\prime \,,
\nonu
D_2 &=& \del^5 + 10 \, u \,\del^3 + 15 \, u^\prime \, \del^2
        + 9 \, u^{\prime\prime} \, \del + 2 \, u^{\prime\prime\prime}
        + 16 \, u^2 \, \del + 16 \, uu^\prime \,.
\label{capd}
\eea

The important observation is now that the differential
equations (\ref{wi}) can be extracted from the Ward Identities defining
the $Sl(3,{\bf R})$ WZW functional, by imposing certain constraints
on the currents \cite{pol2,bo,das,bilal}.
We refer to \cite{us4} for a detailed discussion of this
procedure. As a result of this connection, it is possible to write
an explicit formula for the leading term $\G^{(0)}[h,b]$ in
the induced action. Furthermore, the procedure automatically
leads to a choice of fundamental variables $f_1$ and $f_3$
(instead of $h$ and $b$), which are such that the action
$\G^{(0)}[f_1,f_3]$ takes a local form. The expressions for
$h$ and $b$ in terms of $f_1$ and $f_3$ read
\bea
h &=& \frac{\delb (\frac{\del f_3}{\del f_1})}
{\del (\frac{\del f_3}{\del f_1})} -
\frac{\g}{3} \,
\frac{\del^3 f_3 \del f_1
- \del f_3 \del^3 f_1}{\del^2 f_3 \del f_1 - \del f_3 \del^2 f_1} \, b -
\frac{\g}{2} \, \del b \ ,
\nonu
b &=& \g^{-1} \frac{(\delb f_3  \del f_1  -
\delb f_1 \del f_3 ) } {(\del^2 f_3 \del f_1 -
\del^2 f_1 \del f_3) } \ .
\label{handb}
\eea
The final result for the leading term in the induced action reads
\bea
\lefteqn{ \G^{(0)}[h(f_1,f_3),b(f_1,f_3)] =}
\nonu
&& -\frac{1}{4\p}\int d^2x\,
  \left\{ \frac{\delb\left(\frac{\del f_3}{\del f_1} \right)}
  {\del\left(\frac{\del f_3 }
  {\del f_1}\right)}\del\left(\frac{\del\l_1}{\l_1}\right)
  +\frac{\delb f_1}{\del f_1}\del\left(\frac{\del \l_1}{\l_1}+
  \frac{\del\l_2}{\l_2}\right)\right.
\nonumber\\[3mm]
&& \qquad\qquad\qquad \left. +\frac{\delb f_3\del f_1
-\delb f_1\del f_3}{\del^2f_3\del f_1-\del^2 f_1\del f_3}\left[
\frac{\del\l_2}{\l_2}\del\left(\frac{\del \l_1}{\l_1}\right)-
\frac{\del \l_1}{\l_1}\del\left(\frac{\del\l_2}{\l_2}\right)
\right.\right.
\nonumber\\[3mm]
&& \qquad\qquad\qquad \left.\left.
-\left(\frac{\del\l_1}{\l_1}\right)
\left(\frac{\del\l_2}{\l_2}\right)
\left(\frac{\del\l_1}{\l_1}+
\frac{\del\l_2}{\l_2}\right)\right]\right\} ,
\label{result}
\eea
where
\bea
(\l_1)^3 &=& \left[ \del \left( \frac{\del f_3}{\del f_1} \right)
                   \right]^{-2} (\del f_1)^{-1}
\nonu
(\l_2)^3 &=& \del \left( \frac{\del f_3}{\del f_1} \right) (\del f_1)^{-1}
\label{lambs}
\eea
and $\g^2= - 2/5$.

The effective currents $u$ and $v$ can be written as
\bea
u &=& -\frac 1 2 \left[ \del\left( 2 \frac{\del\l_1}{\l_1}+
\frac{\del\l_2}{\l_2}\right)+
\left(\frac{\del\l_1}{\l_1}\right)^2+
\left(\frac{\del\l_2}{\l_2}\right)^2+
\left(\frac{\del\l_1}{\l_1}\right)
\left(\frac{\del\l_2}{\l_2}\right) \right]
\nonu
v &=& -15\g \left[\frac 1 2 \del^2
\left(\frac{\del\l_2}{\l_2}\right)
+\left(\frac 3 2
\frac{\del\l_1}{\l_1}+\frac{\del \l_2}{\l_2}\right)\del
\left(\frac{\del\l_2}{\l_2}\right)\right.
\nonu
&& \qquad \quad \left. +\frac 1 2
\left(\frac{\del\l_2}{\l_2}\right)
\del\left(\frac{\del\l_1}{\l_1}\right)+
\left(\frac{\del\l_1}{\l_1}+
\frac{\del\l_2}{\l_2}\right)
\left(\frac{\del\l_1}{\l_1}\right)
\left(\frac{\del\l_2}{\l_2}\right)\right].
\label{uandv}
\eea
These expressions, when written in terms of $f_1$ and $f_3$,
can be viewed as `$W_3$ Schwarzian derivatives'. In \cite{us4},
we also discussed an alternative parametrization, which stays
closer to Polyakov's $f$ variable for gravity, and which makes
clear how the truncation from $W_3$ gravity to pure gravity
can be performed.

Let us now turn to the quantization of the $W_3$ gravity fields
$h$ and $b$. The generating functional $W[t,w]$ of connected Green's
functions is defined by
\be
e^{\dis - W[t,w]} = \int {\cal D} h {\cal D} b \,
  e^{\dis - \G_\ind[h,b]+\frac{1}{\p} \int d^2 x \; (h t+ b w) } .
\label{pi2}
\ee

The functional $W[t,w]$ can be analyzed as follows in terms of a $1/c$
expansion for large $c$ (which is the weak coupling regime). We first
approximate the path integral (\ref{pi2}) by the saddle-point contribution.
This leads to the leading term in $W[t,w]$, which is simply the Legendre
transform of the induced action, which by itself was given as a $1/c$
expansion. The saddle-point result should then be corrected by further
terms coming from diagrams with $h$ and $b$ loops. As in the case of
pure gravity, we can see that $1/c$ plays the role of Planck's constant
so that the the loop-corrections to the saddle-point result are
suppressed by strictly positive powers of $1/c$.

In \cite{us5}, we computed the functional $W[t,w]$ through the first
nontrivial order in $1/c$. In doing so, we observed remarkable
cancellations of certain terms, which we view as a sign of the
integrability of this quantum field theory.
The final outcome of our computations through order $1/c$ can be
summarized by the following formula
\bea
&& \!\!\!\!\!
   W[t,w] = \frac{c}{12} \left( 1 - \frac{122}{c} + \ldots \right)
   W_L \left[ \frac{12}{c} \left( 1+\frac{50}{c} + \ldots \right) t,
      \frac{360}{c} \left( 1+\frac{386}{5 c} + \ldots \right) w \right] ,
\nonu &&
\label{result2}
\eea
where the functional $W_L[t,w]$ is related to $\G^{(0)}[h,b]$ by
\be
W_L[u,v] = \G^{(0)}[h(u,v),b(u,v)]
   - \frac{1}{\p} \int d^2 x \; (h u + \frac{1}{30} b v ) \,,
\label{legendre}
\ee
where $h(u,v)$ and $b(u,v)$ are determined through
the relations (\ref{henb}).

We now propose that the exact, all-order result for the functional
$W[t,w]$ can be gotten by simply completing the $1/c$ expansions
indicated by the dots in (\ref{result2}). This leads to the formula
\be
W[t,w] = 2\, k\, W_L \left[ Z^{(t)} t, Z^{(w)} w \right] ,
\label{exact}
\ee
where $k$, $Z^{(t)}$ and $Z^{(w)}$ are functions of $c$ that allow
the $1/c$ expansions
\bea
k &=& \frac{c}{24} \left(1 - \frac{122}{c} + \ldots \right)
\nonu
Z^{(t)} &=& \frac{12}{c} \left( 1 + \frac{50}{c} + \ldots \right)
\nonu
Z^{(w)} &=& \frac{360}{c} \left( 1+ \frac{386}{5 c} + \ldots \right) .
\label{expan}
\eea

The result for $k$ is consistent (in the classical limit
$c \rightarrow -\infty$) with the formula
\be
k = - \frac{1}{48} \left( 50-c+\sqrt{(c-2)(c-98)} \right) -3  \,,
\ee
which is the {\it conjectured}\ outcome of a KPZ type analysis
of constraints in a more covariant formulation of $W_3$ gravity
\cite{bo,matsuo}. Recently, the following all-order results for the
$Z$ factors have been proposed \cite{jandb}:
\be
Z^{(t)}=\frac{1}{2(k+3)} \ , \qquad
Z^{(w)}=\frac{\sqrt{30}}{\sqrt{\b}(k+3)^{3/2}} \ \ .
\ee
They correctly reproduce the singularity structure that one expects,
and are in agreement with the expansions eq.\ (\ref{expan}).
The all-order result for the effective action, which is defined
to be the Legendre transform of $W[t,w]$, follows from (\ref{exact}):
\be
\G_\ef[h,b] = 2\, k \, \G^{(0)} \left[ \frac{1}{2 k Z^{(t)}} h,
  \frac{30}{2 k Z^{(w)}} b \right]  .
\ee

\vspace{6mm}

As we mentioned before, another approach to $W$ gravity,
which avoids studying the gauge sector altogether,
is to use critical theories, {\it i.e.}, theories which are
such that, by a specific choice of the background, all anomalies are made
to cancel. For the purpose of applications in string theory, one would like
to write critical matter systems in terms of scalar `string-coordinate'
fields. The construction of $W$ strings along these lines has
turned out to be much harder than
the construction of ordinary bosonic strings. The reason for this
is not hard to see. The bosonic string requires a $c=26$ contribution to the
central charge from the matter sector for cancellation of the anomaly.
The basic Virasoro multiplet consists of one scalar field. Taking 26 copies
of this theory and coupling them to gravity indeed yields a viable string
theory. The cancellation of the $W_3$ anomalies requires
a matter sector which provides an {\it exact} realization of the $W_3$ algebra
with central charge $c=100$ \cite{brs1}. The basic $W_3$
multiplet consists of two scalar fields ($c=2$). A priori one would expect
that taking 50 copies of this theory would save the day. However,
the resulting theory is not anymore $W_3$ invariant. Its symmetry algebra
contains besides dimension 2 and 3 operators also higher dimensional ones.
One way to obtain matter sectors with $c=100$ is through the introduction
of background charges in the scalar matter sector \cite{hull,romans}.
This was further analyzed in \cite{bgch,bgch2}. However,
the presence of background charges leads to shifts in the mass formulas
\cite{tata}, so that the existence of massless states in the string-spectrum
is in danger (indeed, see \cite{bgch2}). It remains an elusive problem to
find a non-trivial $W_3$ background with $c=100$ and which leads to
massless spin-2 and and possible higher spin states in the string spectrum.

\vspace{6mm}

We would like to thank the organizers of the Summerschool for
giving us the opportunity to present these results.

%%%%%%%%%%%%%%%%%%%%%%%%%%%%%%%%%%%%%%%%%%%%%%%%%%%%%%%%%%%%%%%%%%%%%%%%%%%%%

\end{document}